\documentclass[bibtex,twocolumn,prl,showpacs,floatfix,preprintnumbers,amsmath,amssymb,superscriptaddress,letter]{revtex4}

\usepackage{graphicx}
\usepackage{dcolumn}
\usepackage{color}
\usepackage{xspace}
\usepackage[tight]{units}

\newcommand{\numu}{\ensuremath{\nu_{\mu}}\xspace}
\newcommand{\numubar}{\ensuremath{\overline{\nu}_{\mu}}\xspace}

\newcommand{\nue}{\ensuremath{\nu_{e}}\xspace}

\newcommand{\nuebar}{\ensuremath{\overline{\nu}_{e}}\xspace}
\newcommand{\nutaubar}{\ensuremath{\overline{\nu}_{\tau}}\xspace}

\newcommand{\eV}{\ensuremath{\mathrm{eV}}\xspace}

\newcommand{\trm}[1]{\textrm{#1}}

\newcommand{\dmtwobar}{\ensuremath{|\Delta \overline{m}^{2}|}\xspace}
\newcommand{\dmtwobarnomod}{\ensuremath{\Delta \overline{m}^{2}}\xspace}
\newcommand{\dmtwo}{\ensuremath{|\Delta m^{2}|}\xspace}
\newcommand{\sntwobar}{\ensuremath{\sin^{2}(2\overline{\theta})}\xspace}
\newcommand{\sntwo}{\ensuremath{\sin^{2}(2\theta)}\xspace}

\newcommand{\kNN}{\ensuremath{k\trm{NN}}\xspace}

\newcommand {\comment}[1]{#1\xspace}

\newcommand{\noOscExp}{\ensuremath{156}\xspace}
\newcommand{\observedEvents}{\ensuremath{97}\xspace}
\newcommand{\exposure}{\ensuremath{1.71\times 10^{20}}\xspace}
\newcommand{\confDifference}{\ensuremath{2.0\%}\xspace}
\newcommand{\sigmaToNoOsc}{\ensuremath{6.3}\xspace}
\newcommand{\numuSnNinety}{\ensuremath{0.90}\xspace}
\newcommand{\dmbarBestFitWithErrs}
{\ensuremath{[3.36^{+0.46}_{-0.40}\trm{(stat)}\pm0.06\trm{(syst)}]\times
10^{-3}\,\eV^{2}}\xspace}
\newcommand{\snbarBestFitWithErrs}
{\ensuremath{0.86^{+0.11}_{-0.12}\trm{(stat)}\pm0.01\trm{(syst)}}\xspace}
\newcommand{\dmBestFitWithErrs}
{\ensuremath{(2.32^{+0.12}_{-0.08})\times10^{-3}\,\eV^{2}}\xspace}

\newcommand{\dmBestFit}{\ensuremath{2.32\times10^{-3}\,\eV^{2}}}

\newcommand{\jointdmbestfit}{\ensuremath{2.41\times10^{-3}\,\eV^{2}}}
\newcommand{\jointsnbestfit}{\ensuremath{0.97}}

\begin{document}
\addtolength\topmargin{0.5in}

\preprint{FERMILAB-PUB-11-163-PPD}
\preprint{BNL-94488-2010-JA}
\preprint{arXiv:hep-ex/1104.0344}

\vspace*{1.0cm}

\title{First Direct Observation of Muon Antineutrino Disappearance}

\newcommand{\Berkeley}{Lawrence Berkeley National Laboratory, Berkeley, California, 94720 USA}
\newcommand{\Cambridge}{Cavendish Laboratory, University of Cambridge, Madingley Road, Cambridge CB3 0HE, United Kingdom}
\newcommand{\FNAL}{Fermi National Accelerator Laboratory, Batavia, Illinois 60510, USA}
\newcommand{\RAL}{Rutherford Appleton Laboratory, Science and Technologies Facilities Council, OX11 0QX, United Kingdom}
\newcommand{\UCL}{Department of Physics and Astronomy, University College London, Gower Street, London WC1E 6BT, United Kingdom}
\newcommand{\Caltech}{Lauritsen Laboratory, California Institute of Technology, Pasadena, California 91125, USA}
\newcommand{\Alabama}{Department of Physics and Astronomy, University of Alabama, Tuscaloosa, Alabama 35487, USA}
\newcommand{\ANL}{Argonne National Laboratory, Argonne, Illinois 60439, USA}
\newcommand{\Athens}{Department of Physics, University of Athens, GR-15771 Athens, Greece}
\newcommand{\NTUAthens}{Department of Physics, National Tech. University of Athens, GR-15780 Athens, Greece}
\newcommand{\Benedictine}{Physics Department, Benedictine University, Lisle, Illinois 60532, USA}
\newcommand{\BNL}{Brookhaven National Laboratory, Upton, New York 11973, USA}
\newcommand{\CdF}{APC -- Universit\'{e} Paris 7 Denis Diderot, 10, rue Alice Domon et L\'{e}onie Duquet, F-75205 Paris Cedex 13, France}
\newcommand{\Cleveland}{Cleveland Clinic, Cleveland, Ohio 44195, USA}
\newcommand{\Delhi}{Department of Physics \& Astrophysics, University of Delhi, Delhi 110007, India}
\newcommand{\GEHealth}{GE Healthcare, Florence South Carolina 29501, USA}
\newcommand{\Harvard}{Department of Physics, Harvard University, Cambridge, Massachusetts 02138, USA}
\newcommand{\HolyCross}{Holy Cross College, Notre Dame, Indiana 46556, USA}
\newcommand{\IIT}{Physics Division, Illinois Institute of Technology, Chicago, Illinois 60616, USA}
\newcommand{\Iowa}{Department of Physics and Astronomy, Iowa State University, Ames, Iowa 50011 USA}
\newcommand{\Indiana}{Indiana University, Bloomington, Indiana 47405, USA}
\newcommand{\ITEP}{High Energy Experimental Physics Department, ITEP, B. Cheremushkinskaya, 25, 117218 Moscow, Russia}
\newcommand{\JMU}{Physics Department, James Madison University, Harrisonburg, Virginia 22807, USA}
\newcommand{\LASL}{Nuclear Nonproliferation Division, Threat Reduction Directorate, Los Alamos National Laboratory, Los Alamos, New Mexico 87545, USA}
\newcommand{\Lebedev}{Nuclear Physics Department, Lebedev Physical Institute, Leninsky Prospect 53, 119991 Moscow, Russia}
\newcommand{\LLL}{Lawrence Livermore National Laboratory, Livermore, California 94550, USA}
\newcommand{\LosAlamos}{Los Alamos National Laboratory, Los Alamos, New Mexico 87545, USA}
\newcommand{\MIT}{Lincoln Laboratory, Massachusetts Institute of Technology, Lexington, Massachusetts 02420, USA}
\newcommand{\Minnesota}{University of Minnesota, Minneapolis, Minnesota 55455, USA}
\newcommand{\Crookston}{Math, Science and Technology Department, University of Minnesota -- Crookston, Crookston, Minnesota 56716, USA}
\newcommand{\Duluth}{Department of Physics, University of Minnesota -- Duluth, Duluth, Minnesota 55812, USA}
\newcommand{\Ohio}{Center for Cosmology and Astro Particle Physics, Ohio State University, Columbus, Ohio 43210 USA}
\newcommand{\Otterbein}{Otterbein College, Westerville, Ohio 43081, USA}
\newcommand{\Oxford}{Subdepartment of Particle Physics, University of Oxford, Oxford OX1 3RH, United Kingdom}
\newcommand{\PennState}{Department of Physics, Pennsylvania State University, State College, Pennsylvania 16802, USA}
\newcommand{\PennU}{Department of Physics and Astronomy, University of Pennsylvania, Philadelphia, Pennsylvania 19104, USA}
\newcommand{\Pittsburgh}{Department of Physics and Astronomy, University of Pittsburgh, Pittsburgh, Pennsylvania 15260, USA}
\newcommand{\IHEP}{Institute for High Energy Physics, Protvino, Moscow Region RU-140284, Russia}
\newcommand{\Rochester}{Department of Physics and Astronomy, University of Rochester, New York 14627 USA}
\newcommand{\RoyalH}{Physics Department, Royal Holloway, University of London, Egham, Surrey, TW20 0EX, United Kingdom}
\newcommand{\Carolina}{Department of Physics and Astronomy, University of South Carolina, Columbia, South Carolina 29208, USA}
\newcommand{\SLAC}{Stanford Linear Accelerator Center, Stanford, California 94309, USA}
\newcommand{\Stanford}{Department of Physics, Stanford University, Stanford, California 94305, USA}
\newcommand{\StJohnFisher}{Physics Department, St. John Fisher College, Rochester, New York 14618 USA}
\newcommand{\Sussex}{Department of Physics and Astronomy, University of Sussex, Falmer, Brighton BN1 9QH, United Kingdom}
\newcommand{\TexasAM}{Physics Department, Texas A\&M University, College Station, Texas 77843, USA}
\newcommand{\Texas}{Department of Physics, University of Texas at Austin, 1 University Station C1600, Austin, Texas 78712, USA}
\newcommand{\TechX}{Tech-X Corporation, Boulder, Colorado 80303, USA}
\newcommand{\Tufts}{Physics Department, Tufts University, Medford, Massachusetts 02155, USA}
\newcommand{\UNICAMP}{Universidade Estadual de Campinas, IFGW-UNICAMP, CP 6165, 13083-970, Campinas, SP, Brazil}
\newcommand{\UFG}{Instituto de F\'{i}sica, Universidade Federal de Goi\'{a}s, CP 131, 74001-970, Goi\^{a}nia, GO, Brazil}
\newcommand{\USP}{Instituto de F\'{i}sica, Universidade de S\~{a}o Paulo,  CP 66318, 05315-970, S\~{a}o Paulo, SP, Brazil}
\newcommand{\Warsaw}{Department of Physics, Warsaw University, Ho\.{z}a 69, PL-00-681 Warsaw, Poland}
\newcommand{\Washington}{Physics Department, Western Washington University, Bellingham, Washington 98225, USA}
\newcommand{\WandM}{Department of Physics, College of William \& Mary, Williamsburg, Virginia 23187, USA}
\newcommand{\Wisconsin}{Physics Department, University of Wisconsin, Madison, Wisconsin 53706, USA}
\newcommand{\deceased}{Deceased.}

\affiliation{\ANL}
\affiliation{\Athens}
\affiliation{\BNL}
\affiliation{\Caltech}
\affiliation{\Cambridge}
\affiliation{\UNICAMP}
\affiliation{\FNAL}
\affiliation{\UFG}
\affiliation{\Harvard}
\affiliation{\HolyCross}
\affiliation{\IIT}
\affiliation{\Indiana}
\affiliation{\Iowa}
\affiliation{\UCL}
\affiliation{\Minnesota}
\affiliation{\Duluth}
\affiliation{\Otterbein}
\affiliation{\Oxford}
\affiliation{\Pittsburgh}
\affiliation{\RAL}
\affiliation{\USP}
\affiliation{\Carolina}
\affiliation{\Stanford}
\affiliation{\Sussex}
\affiliation{\TexasAM}
\affiliation{\Texas}
\affiliation{\Tufts}
\affiliation{\Warsaw}
\affiliation{\WandM}

\author{P.~Adamson}
\affiliation{\FNAL}

\author{C.~Andreopoulos}
\affiliation{\RAL}



\author{D.~J.~Auty}
\affiliation{\Sussex}


\author{D.~S.~Ayres}
\affiliation{\ANL}

\author{C.~Backhouse}
\affiliation{\Oxford}




\author{G.~Barr}
\affiliation{\Oxford}









\author{M.~Bishai}
\affiliation{\BNL}

\author{A.~Blake}
\affiliation{\Cambridge}


\author{G.~J.~Bock}
\affiliation{\FNAL}

\author{D.~J.~Boehnlein}
\affiliation{\FNAL}

% ...

\collaboration{MINOS Collaboration}
\noaffiliation

\begin{abstract}

This Letter reports the first direct observation of muon antineutrino
disappearance. The MINOS experiment has taken data with an accelerator
beam optimized for \numubar production, accumulating an exposure of
\exposure protons on target. In the Far Detector, \observedEvents
charged current \numubar events are observed. The no-oscillation
hypothesis predicts \noOscExp events and is excluded at
$\sigmaToNoOsc\sigma$.  The best fit to oscillation yields
$\dmtwobar=\dmbarBestFitWithErrs$, $\sntwobar=\snbarBestFitWithErrs$.
The MINOS \numu and \numubar measurements are consistent at the
\confDifference confidence level, assuming identical underlying
oscillation parameters.
\end{abstract}
\maketitle


Observations by many experiments provide compelling evidence for
neutrino \mbox{oscillation
\cite{ref:minos2006,ref:minos2008,ref:minosCC2010,ref:sk,ref:soudan2,ref:macro,ref:k2k,ref:sno,ref:kamland}}. This
oscillation, a consequence of the quantum mechanical mixing of the
neutrino mass and weak flavor eigenstates, is governed by the elements
of the Pontecorvo-Maki-Nakagawa-Sakata \mbox{matrix \cite{ref:pmns}},
parameterized by three mixing angles and a {\it CP} phase, and by two
independent neutrino mass-squared differences.  As the measurement
precision on oscillation parameters improves, so does the potential
for observing new phenomena. In particular, measured differences
between the neutrino and antineutrino oscillation parameters would
indicate new physics.  {\it CPT} symmetry, one of the most fundamental
assumptions underlying the standard model, constrains the allowed
differences in the properties of a particle and its antiparticle
\cite{ref:weinberg} and requires that their masses be identical. This
symmetry has been extensively tested in other sectors, most notably
the kaon sector~\cite{ref:pdg}. Additionally, neutrinos passing
through matter could experience nonstandard \mbox{interactions
\cite{ref:NSI}} that alter the \numu and \numubar disappearance
probabilities and, thus, the inferred oscillation
parameters~\cite{ref:minosNSIfits}.

The MINOS experiment has used a \numu beam to measure the larger
(atmospheric) mass-squared difference $\dmtwo=\dmBestFitWithErrs$
and the mixing angle $\sntwo>\numuSnNinety$ (90\% confidence limit
[C.L.])  through observation of \numu \mbox{disappearance
  \cite{comment:dmbar,ref:minosCC2010}}.
The corresponding antineutrino oscillation parameters are much less
precisely known. 

This Letter describes the first direct observation of \numubar
disappearance consistent with oscillation, yielding the most precise
measurement to date of the larger antineutrino mass-squared
difference. The only previous measurements from \numubar-tagged
samples, providing weak constraints, come from the MINOS atmospheric
neutrino \mbox{sample \cite{ref:minosAtmos}} and an analysis of the
\numubar component of the MINOS \numu data
sample~\cite{ref:JustinThesis, ref:AutyThesis}.  The strongest
indirect constraints come from a global \mbox{fit \cite{ref:maltoni}},
dominated by Super-Kamiokande data which measure the sum of
atmospheric \numu and \numubar interaction rates.

\begin{figure}
\centering
\includegraphics[width=\columnwidth]{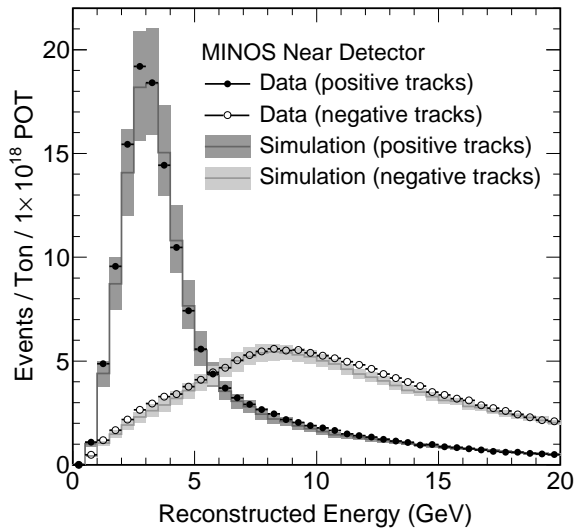}
\caption{The reconstructed energy spectra of events in the Near
Detector classified as charged current interactions, separated
according to the reconstructed charge of the track. The events with a
negatively charged track are not used in the oscillation analysis. The
shaded bands represent the systematic uncertainty on the simulation.}
\label{fig:NDEnergy}
\end{figure}

For this measurement the NuMI \mbox{beam line \cite{ref:numi}} was
configured to produce a \numubar-enhanced beam. The current in the
magnetic horns was configured to focus negative pions and kaons
produced by \unit[120]{GeV} protons incident on a graphite
target. Most mesons travel along a \unit[675]{m} long decay pipe,
filled with helium at \unit[0.9]{atm}, and decay to produce a
\numubar-enhanced beam with a peak energy of \unit[3]{GeV} (see
\mbox{Fig.\ \ref{fig:NDEnergy}}). Interactions of \numu comprise a
fraction of all charged current (CC) events in the MINOS detectors
which rises from about 21\% below \unit[6]{GeV} up to about 81\% at
\unit[20]{GeV}, in the case of no oscillation.
The data set in this paper corresponds to an exposure of
\exposure protons on target (POT).

The MINOS experiment uses two similar detectors located
\unit[1.04]{km} [Near Detector (ND)] and \unit[735]{km} [Far Detector
(FD)] from the NuMI target. The \numubar CC interaction rate as a
function of reconstructed \numubar energy is measured in each
detector. The measured FD energy spectrum is compared to that
predicted using the ND data. In this comparison, many sources of
systematic uncertainty cancel. Antineutrino oscillation causes a
deficit in the FD with an energy dependence, in the approximation of
two-flavor mixing, of
\begin{equation}
P(\numubar\rightarrow\numubar) =
1 - \sntwobar\sin^{2}\left(
\frac{1.267\dmtwobarnomod L}{E}
\right)
\label{eq:osc}
\end{equation}
where $L$ [km] is the distance from the point of antineutrino
production, $E$ [GeV] the \numubar energy, $\dmtwobarnomod$
[$\eV^{2}$] the antineutrino mass-squared difference and
$\overline{\theta}$ the antineutrino mixing angle.

The MINOS \mbox{detectors \cite{ref:minosNim}} are tracking
calorimeters, formed of planes of steel interleaved with planes of
scintillator. The scintillator is divided into strips with a width of
\unit[4.1]{cm}. In CC interactions,
$\numubar(\numu)+N\rightarrow\mu^{+}(\mu^{-})+X$, a hadronic shower
($X$) and a muon track may be observed. The hadronic energy is
measured by summing the amount of light produced in the
scintillator. Muon energy is measured by range for contained tracks
or, for exiting tracks, by the curvature in a \unit[$\sim 1.4$]{T}
toroidal magnetic field. The incoming neutrino energy is reconstructed
as the sum of the hadronic and muon energies. For the data presented
in this Letter, the fields in both detectors focus $\mu^{+}$ and
defocus $\mu^{-}$, allowing the separation of \numubar and \numu CC
interactions on an event-by-event basis.

A sample of \numubar CC interactions is isolated by identifying the
presence of a positively charged track. Neutral current (NC)
interactions produce only a hadronic shower at the vertex. Similarly,
CC interactions of \nue and \nuebar (which correspond to 2.0\% of all
CC interactions at the ND) produce only showerlike activity. The main
background arises from tracks reconstructed out of shower
activity. This background is reduced
\cite{ref:minos2008,ref:RustemThesis} by a method which uses four
variables to identify the presence of an isolated track with muonlike
energy deposition. These four variables are the track length, the
average pulse height per plane along the track, the transverse energy
deposition profile of the track, and the fluctuation of the energy
deposited in scintillator strips along the track, and are combined
using a $k$-nearest-neighbor (\kNN) \mbox{algorithm \cite{ref:kNN}} to
produce a single output variable. The position of the selection cut on
this variable is tuned to optimize the statistical sensitivity to
$\dmtwobar$, yielding the same selection criterion as for the MINOS
\numu analysis~\cite{ref:minos2008}.

The charge of reconstructed muon tracks is determined by analyzing the
curvature of the track in the magnetic \mbox{field
\cite{ref:JohnMarshallThesis}}. Figure~\ref{fig:NDEnergy} shows the
reconstructed energy of selected CC events in the ND, separated
according to the measured track charge sign.  The events reconstructed
with a negatively charged track consist primarily of \numu CC
interactions, and are removed from further analysis. Events with a
positively charged track form the selected \numubar CC sample, and are
used to predict the expected energy spectrum at the FD. Below
\unit[6]{GeV}, where the majority of the oscillation signal is
expected, the selected \numubar CC sample at the ND has a purity,
obtained from the simulation, of 98\% (the background consisting of
1\% NC events, 1\% \numu CC events). Above \unit[6]{GeV} the purity is
88\%, and the contamination is primarily \numu CC events; higher
momentum muons follow a less curved path, giving a greater probability
of charge misidentification. The total \numubar CC reconstruction and
selection efficiency is $93\%$.

The measured ND energy spectrum is used to predict the FD spectrum, as
previously~\cite{ref:minos2006,ref:minos2008,ref:JustinThesis}. This
procedure is particularly effective in mitigating sources of
systematic uncertainty which affect both detectors similarly. For
example, uncertainties on the neutrino flux and cross sections
dominate the systematic error band on the ND energy spectrum, shown in
\mbox{Fig.\ \ref{fig:NDEnergy}}, but have a negligible impact on the
oscillation measurement.

The production of hadrons in the NuMI target is constrained by fits to
the ND
data~\cite{ref:minos2006,ref:minos2008}. These
fits use data from the \numu beam to determine the $\pi$ and $K$
yields as a function of their transverse and longitudinal momenta at
production. Recent measurements~\cite{ref:na49} of the ratio of
$\pi^{+}/\pi^{-}$ yields are included as constraints in these
fits. This tuning procedure improves agreement between the simulated
ND energy spectrum and the data, but does not significantly affect the
predicted FD energy spectrum. Uncertainties on the modeling of the
beam have a negligible effect on the predicted FD energy spectrum, and
are accounted for in the oscillation measurement.

\begin{figure}
\centering
\includegraphics[width=\columnwidth]{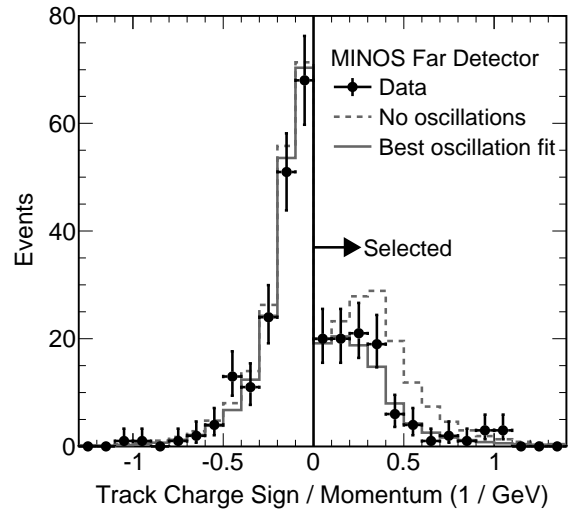}
\caption{The distribution of the sign of the reconstructed charge
divided by the momentum of selected muon tracks in the Far
Detector. The simulated distribution is shown in the case of no
oscillation and oscillation assuming the best fit \numu parameters
from Ref.~\cite{ref:minosCC2010} and \numubar parameters from this
analysis.}
\label{fig:FDqp}
\end{figure}

The same event selection criteria are used in both detectors. The FD
data selection was determined using simulation and ND data, before the
FD data was examined. All FD events passing the \kNN selection are
shown in \mbox{Fig.\ \ref{fig:FDqp}}, distributed according to the
sign of the reconstructed track charge, divided by the momentum. The
figure shows good modeling of track charge identification. Events with
a negatively charged track are minimally affected by oscillation due
to their higher mean energy, and are removed from further analysis.

The systematic uncertainty on the oscillation parameters is much
smaller than the statistical uncertainty. The sources of systematic
uncertainty are very similar to those discussed for the MINOS \numu
analysis~\cite{ref:minosCC2010}. An additional uncertainty is
estimated on the level of \numu CC background in the selected \numubar
CC event sample; below \unit[6]{GeV}, the purity of the selected
\numubar CC sample is known to within $1\%$. To evaluate this
uncertainty, the simulated background is scaled until the total number
of simulated and data events match in the background-enhanced set of
events which fail the \kNN selection criterion. This scale factor is
taken as the uncertainty on the level of background in the selected
\numubar CC sample. The total systematic uncertainty on the
measurement of \dmtwobar is
\unit[$+0.063-0.060\times10^{-3}$]{eV$^{2}$}; on the
measurement of \sntwobar the total systematic uncertainty is
$\pm0.012$.

\begin{figure}
\centering
\includegraphics[width=\columnwidth]{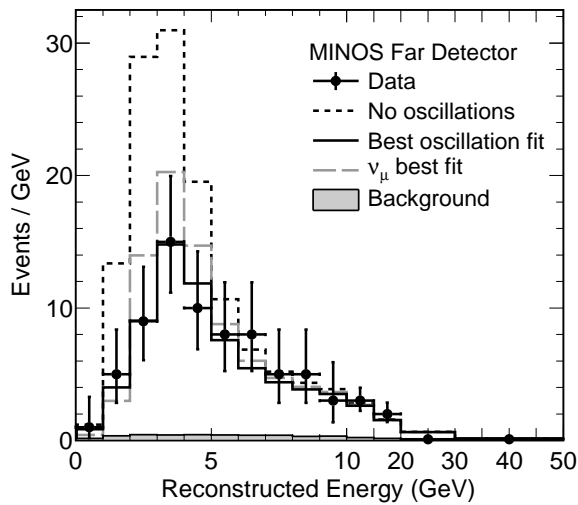}
\caption{Comparison of the measured Far Detector \numubar CC energy
spectrum to the expectation in three cases: in the absence of
oscillation, using the oscillation parameters which best fit this
\numubar data (for this case, the total expected background is also
indicated), and using the best-fit \numu oscillation parameters
measured by \mbox{MINOS \cite{ref:minosCC2010}}.}
\label{fig:FDspectrum}
\end{figure}

Using the prediction obtained from the ND data, \noOscExp selected
\numubar CC events with energy below \unit[50]{GeV} are expected in
the FD in the absence of oscillation while \observedEvents events are
observed. The energy spectra of these FD events are shown in
\mbox{Fig.\ \ref{fig:FDspectrum}}. A clear energy dependent deficit is
observed, showing the first conclusive evidence for \numubar
disappearance consistent with oscillation in a \numubar-tagged
sample. The no-oscillation hypothesis is disfavored at \sigmaToNoOsc
standard deviations.

Oscillation is incorporated into the predicted energy spectrum
according to \mbox{Eq.\ \ref{eq:osc}}. Comparing the prediction to the
data using a binned log likelihood, oscillation parameters are found
which maximize the likelihood. These are
$\dmtwobar=\dmbarBestFitWithErrs$ and
$\sntwobar=\snbarBestFitWithErrs$, and are consistent with all other
previous direct
limits~\cite{ref:minosAtmos,ref:JustinThesis,ref:AutyThesis}. The
prediction for oscillation with these best fit values is shown in
\mbox{Fig.\ \ref{fig:FDspectrum}}, and includes 2 NC events, 5 \numu
CC events and 0.3 \nutaubar CC events.

\begin{figure}
\centering
\includegraphics[width=\columnwidth]{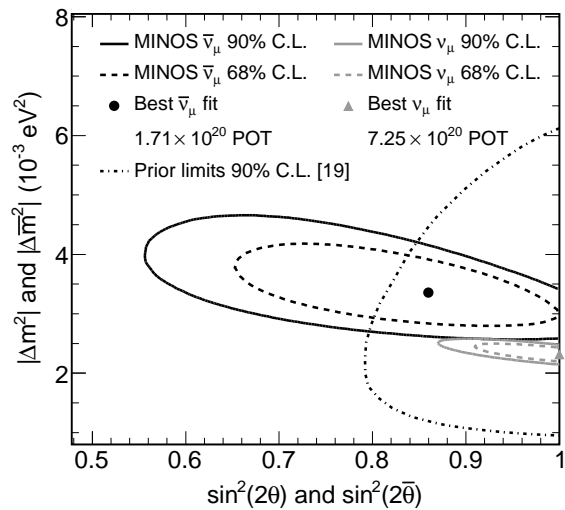}
\caption{Allowed regions for the \numubar oscillation parameters from
a fit to the data in \mbox{Fig.\ \ref{fig:FDspectrum}}, including all
sources of systematic uncertainty. Indirect limits prior to this
\mbox{work \cite{ref:maltoni}} and the MINOS allowed region for \numu
\mbox{oscillation \cite{ref:minosCC2010}} are also shown.}
\label{fig:Contours}
\end{figure}

The confidence limits on the oscillation parameters, shown in
Fig.~\ref{fig:Contours}, are calculated using the Feldman-Cousins
technique~\cite{ref:FeldCous}, in which the effect of all sources of
systematic uncertainty is
included~\cite{ref:AlexThesis,ref:NickThesis}. Figure~\ref{fig:Contours}
compares these limits to the previous best \mbox{limit
\cite{ref:maltoni}}.

MINOS has measured the \numu oscillation parameters to obtain a best
fit of $\dmtwo=\dmBestFit$, \mbox{$\sntwo=1.0$
\cite{ref:minosCC2010}}. Assuming that muon antineutrinos oscillate
with these parameters, 110 selected events are expected in the FD
below \unit[50]{GeV}. This expected energy spectrum is shown in
Fig.~\ref{fig:FDspectrum}, denoted as ``\numu best fit''.

\mbox{Figure \ref{fig:Contours}} compares the MINOS measurements of
the \numu and \numubar oscillation parameters. In both measurements
the purity of the event samples in the oscillation region is
high. Below \unit[6]{GeV}, there is no more than 3\% \numu CC
contamination in the \numubar CC sample and vice-versa. Therefore the
measurements of the \numu and \numubar oscillation parameters are
nearly independent. Since the \numubar measurement is heavily
statistically limited, the impact of correlated systematic
uncertainties is negligible.

In a joint fit to the data used in the MINOS \numu and \numubar
measurements, assuming identical \numu and \numubar oscillation
parameters, the best fit parameters are
$\dmtwobar\equiv\dmtwo=\jointdmbestfit$,
$\sntwobar\equiv\sntwo=\jointsnbestfit$. The significance of the
difference in likelihood between this best fit and the fits to the
individual \numu and \numubar data sets is evaluated using a
Feldman-Cousins approach~\cite{ref:AlexThesis}. Ten thousand simulated
experiments are generated assuming the joint best fit oscillation
parameters above, and include all sources of systematic
uncertainty. The difference in likelihood between the joint and
individual \numu and \numubar fits is recorded for each experiment,
and the fraction of simulated experiments with a difference in
likelihood larger than that observed in the data is a measure of the
significance of the observed difference.  Assuming identical \numu and
\numubar oscillation parameters, the probability that the MINOS
measurements of the \numu and \numubar parameters would be more
discrepant than those observed is \confDifference.

A thorough search for sources of mismodeling that could provide a
difference between the \numu and \numubar measurements was performed.
The only change between \numu and \numubar running modes was the
reversal of the directions of the current in the focusing horns of the
beam and of the magnetic fields in the detectors. Very similar data
analysis procedures are used in both modes, with the same
reconstruction code and similar selection criteria used to obtain the
event samples, and the same technique used to obtain the FD
predictions. These similarities make the comparison of \numu and
\numubar measurements robust and limit the possible sources which
could generate a spurious difference.

The \numu and \numubar analyses differ in that the \numubar-enhanced
beam contains a significant \numu component (which is effectively
removed by the selection cuts). \mbox{Figure \ref{fig:NDEnergy}} shows
that this component is at high energy, away from the oscillation
signal region, and therefore any residual contamination has little
effect on the oscillation measurement. \mbox{Figures
\ref{fig:NDEnergy}} \mbox{and \ref{fig:FDqp}} show the \numu CC
component to be well modeled in both detectors in both shape and
normalization. All FD events were scanned by eye to ensure the
selection does not anomalously mis-classify events by the sign of the
charge. Performing the analysis without any selection on track charge
in the FD produces consistent results.  A scan by eye of events in the
ND showed the track reconstruction efficiency to be well modeled.

Differences in the underlying inelasticity distributions for \numu and
\numubar events lead to differences in the muon and hadron energy
distributions for the two samples.  Studies to validate the
reconstruction of muon tracks and hadronic showers included the
tightening of reconstruction quality criteria, investigation of the
comparative performance of various detector regions, and the use of an
alternative hadronic energy estimator. These studies show the
detectors to be well modeled, and that any mismodeling in
reconstruction and selection efficiencies is accounted for in the
systematic uncertainty. In conclusion, no evidence is found for any
systematic error that could cause a significant difference between the
measured \numu and \numubar oscillation parameters.

In summary, MINOS has used a beam optimized for the production of
\numubar to make the first direct observation of \numubar
disappearance and, assuming the disappearance is caused by
oscillation, has made the most precise measurement of the
corresponding antineutrino mass-squared difference to date. From fits
to these data, MINOS has measured the oscillation parameters to be
$\dmtwobar=\dmbarBestFitWithErrs$ and
$\sntwobar=\snbarBestFitWithErrs$.  The MINOS \numu and \numubar
measurements are consistent at the \confDifference confidence level,
assuming identical underlying oscillation parameters.
Additional data are currently being taken with the
\numubar-enhanced NuMI beam, with the aim of doubling the statistics
presented in this Letter.

This work was supported by the U.S. DOE; the United Kingdom STFC; the
U.S. NSF; the State and University of Minnesota; the University of
Athens, Greece; and Brazil's FAPESP and CNPq.  We are grateful to the
Minnesota Department of Natural Resources, the crew of the Soudan
Underground Laboratory, and the personnel of Fermilab for their
contributions to this effort.



\begin{thebibliography}{99}
\bibitem{comment:dmbar} The experiment measures an unresolved mixture
of $|\Delta m^{2}_{31}|$ and $|\Delta
m^{2}_{32}|$, which is referred to as $\dmtwo$ for
brevity. The parameter \sntwo is likewise an admixture, dominated
by $\theta_{23}$. Similarly for \dmtwobar and \sntwobar.
\end{thebibliography}
\end{document}